\documentclass[aps, prb,reprint,superscriptaddress,showpacs, citeautoscript]{revtex4-1}

\usepackage{graphicx}
\usepackage{dcolumn}
\usepackage{bm}
\usepackage{natbib}
\usepackage{amsfonts}
\usepackage{amsmath}
\usepackage{amssymb}

\begin{document}

\title{Magnetic excitations and phonons simultaneously studied by resonant inelastic
x-ray scattering in optimally doped
Bi$_{1.5}$Pb$_{0.55}$Sr$_{1.6}$La$_{0.4}$CuO$_{6+\delta}$}

\author{Y.~Y.~Peng}
\email{yingying.peng@polimi.it} \affiliation{Dipartimento di Fisica,
Politecnico di Milano, Piazza Leonardo da Vinci 32, I-20133 Milano, Italy}
\author{M.~Hashimoto}
\affiliation{Stanford Synchrotron Radiation Lightsource, SLAC National
Accelerator Laboratory, 2575, Sand Hill Road, Menlo Park, California 94025,
USA}
\author{M.~Moretti Sala}
\affiliation{European Synchrotron Radiation Facility (ESRF), BP 220, F-38043
Grenoble Cedex, France}
\author{A.~Amorese}
\altaffiliation[Present address: ]{European Synchrotron Radiation Facility
(ESRF), BP 220, F-38043 Grenoble Cedex, France} \affiliation{Dipartimento di
Fisica, Politecnico di Milano, Piazza Leonardo da Vinci 32, I-20133 Milano,
Italy}
\author{N.~B.~Brookes}
\affiliation{European Synchrotron Radiation Facility (ESRF), BP 220, F-38043
Grenoble Cedex, France}
\author{G.~Dellea}
\affiliation{Dipartimento di Fisica, Politecnico di Milano, Piazza Leonardo da
Vinci 32, I-20133 Milano, Italy}
\author{W.-S.~Lee}
\affiliation{Stanford Institute for Materials and Energy Sciences, SLAC
National Accelerator Laboratory, 2575 Sand Hill Road, Menlo Park, CA 94025,
USA}
\author{M.~Minola}
\altaffiliation[Present address: ]{Max-Planck-Institut f\"{u}r
Festk\"{o}rperforschung, Heisenbergstra{\ss}e 1, D-70569 Stuttgart, Germany}
\affiliation{Dipartimento di Fisica, Politecnico di Milano, Piazza Leonardo da
Vinci 32, I-20133 Milano, Italy}
\author{T.~Schmitt}
\affiliation{Swiss Light Source, Paul Scherrer Institut, CH-5232 Villigen PSI,
Switzerland}
\author{Y.~Yoshida}
\affiliation{Nanoelectronics Research Institute, AIST, Ibaraki 305-8568, Japan}
\author{K.-J.~Zhou}
\altaffiliation[Present address: ]{Diamond Light Source, Harwell Science and
Innovation Campus, Didcot, Oxon, OX11 0DE, United Kingdom} \affiliation{Swiss
Light Source, Paul Scherrer Institut, CH-5232 Villigen PSI, Switzerland}
\author{H.~Eisaki}
\affiliation{Nanoelectronics Research Institute, AIST, Ibaraki 305-8568, Japan}
\author{T.~P.~Devereaux}
\affiliation{Stanford Institute for Materials and Energy Sciences, SLAC
National Accelerator Laboratory, 2575 Sand Hill Road, Menlo Park, CA 94025,
USA}
\author{Z.-X.~Shen}
\affiliation{Stanford Institute for Materials and Energy Sciences, SLAC
National Accelerator Laboratory, 2575 Sand Hill Road, Menlo Park, CA 94025,
USA} \affiliation{Geballe Laboratory for Advanced Materials, Departments of
Physics and Applied Physics, Stanford University, CA 94305, USA}
\author{L.~Braicovich}
\affiliation{Dipartimento di Fisica, Politecnico di Milano, Piazza Leonardo da
Vinci 32, I-20133 Milano, Italy} \affiliation{CNR-SPIN, CNISM, Politecnico di
Milano, Piazza Leonardo da Vinci 32, I-20133 Milano, Italy}
\author{G.~Ghiringhelli}
\affiliation{Dipartimento di Fisica, Politecnico di Milano, Piazza Leonardo da
Vinci 32, I-20133 Milano, Italy} \affiliation{CNR-SPIN, CNISM, Politecnico di
Milano, Piazza Leonardo da Vinci 32, I-20133 Milano, Italy}

\date{\today}

\begin{abstract}

Magnetic excitations in the optimally doped high-$T_\mathrm{c}$ superconductor
Bi$_{1.5}$Pb$_{0.55}$Sr$_{1.6}$La$_{0.4}$CuO$_{6+\delta}$ (OP-Bi2201,
$T_\mathrm{c}\simeq 34$ K) are investigated by Cu $L_3$ edge resonant inelastic
x-ray scattering (RIXS), below and above the pseudogap opening temperature. At
both temperatures the broad spectral distribution disperses along the (1,0)
direction up to $\sim$350~meV at zone boundary, similarly to other hole-doped
cuprates. However, above $\sim$0.22 reciprocal lattice units, we observe a
concurrent intensity decrease for magnetic excitations and quasi-elastic
signals with weak temperature dependence. This anomaly seems to indicate a
coupling between magnetic, lattice and charge modes in this compound. We also
compare the magnetic excitation spectra near the anti-nodal zone boundary in
the single layer OP-Bi2201 and in the bi-layer optimally doped
Bi$_{1.5}$Pb$_{0.6}$Sr$_{1.54}$CaCu$_2$O$_{8+\delta}$ (OP-Bi2212,
$T_\mathrm{c}\simeq96$ K). The strong similarities in the paramagnon dispersion
and in their energy at zone boundary indicate that the strength of the
super-exchange interaction and the short-range magnetic correlation cannot be
directly related to $T_\mathrm{c}$, not even within the same family of
cuprates.

\end{abstract}


\maketitle

\section{Introduction}

In recent years, resonant inelastic x-ray scattering (RIXS) at the Cu $L_3$
edge, thanks to the strong spin-orbit coupling of the 2$p_{3/2}$ core-hole
intermediate state that provides a direct access to spin flip excitations
\cite{AmentMagnon,HaverkortLS}, has become a powerful complement to neutron
inelastic scattering for the determination of magnetic excitation dispersion in
cuprates. The persistence of magnetic excitations has been observed from
undoped antiferromagnetic insulators to overdoped superconductors, both for
hole- and electron-doped compounds, across the respective superconducting domes
\cite{LucioLSCO,TaconNatphysics,DeanNatMat,DeanPRL,TaconPRBOD,IshiElectrondope}.
This confirms that short-range antiferromagnetic spin correlations survive to
high doping and are exceptionally robust. These observations might imply that
high energy magnetic fluctuations are necessary for superconductivity, although
their possible role as pairing mechanism has not been demonstrated.

Moreover Cu $L_3$ resonant soft x-ray scattering has decisively contributed to
reveal an electronic order now considered ubiquitous in the cuprate
superconductors. Earlier evidence of bulk charge order was observed in La-based
cuprates by neutron \cite{TranquadaStripe,FujitaStripe} and x-ray scattering
\cite{Stripexray}. More recently, charge order along the Cu-O bond direction
has been observed in (Y,Nd)Ba$_2$Cu$_3$O$_{6+x}$ (YBCO,
NBCO)\cite{JulienNMR,GiacomoCDW,REXSCDW,changCDW,diffractionCDW}, Bi-based
cuprates Bi$_2$Sr$_{2-x}$La$_x$CuO$_{6+\delta}$ (Bi2201) and
Bi$_2$Sr$_2$CaCu$_2$O$_{8+\delta}$ (Bi2212)
\cite{Damascelliscience,STMBi2212,HashimotoBi2212}. The direct observation of
charge order in cuprates came first in the underdoped regime, soon after for
the optimal doping of hole-doped cuprates\cite{HashimotoBi2212,TaconYBCOarxiv}
and very recently also in the electron-doped compounds \cite{CominNBCO}. The
temperature dependence of charge order in YBCO and Bi2212 displays its
competition with superconductivity \cite{GiacomoCDW,changCDW,STMBi2212}.

To comprehend the superconductivity of cuprates we need to take into account
the diversity of electronic and magnetic excitations. Whether and how these
excitations, such as charge, spin, lattice and orbital orders, interact with
each other is still the matter of active research. One of the superiority of
RIXS is that it can measure different kinds of excitations simultaneously. This
allows us to discuss the interplay between the charge, spin, lattice and
orbital orders more in detail within a single experiment. Further, temperature
dependence may provide insights into their physical properties and about their
role in superconductivity and pseudogap.

In this Article we study the collective excitations in the optimally doped
Bi$_{1.5}$Pb$_{0.55}$Sr$_{1.6}$La$_{0.4}$CuO$_{6+\delta}$ (OP-Bi2201,
$T_\textrm{c} \simeq 34$ K) by momentum resolved resonant inelastic x-ray
scattering (RIXS) at the Cu $L_3$ edge. Considering that the recent
angle-resolved photoemission spectroscopy (ARPES) study on OP-Bi2201 showed a
particle-hole symmetry breaking and a phase transition below the pseudogap
temperature $T^* \simeq 125$ K) \cite{Hashimotonphys, Hescience}, our
measurements were performed at two temperatures, 50~K (below $T^*$) and 200~K
(above $T^*$). We also compare these results to those of optimally doped
Bi$_{1.5}$Pb$_{0.6}$Sr$_{1.54}$CaCu$_2$O$_{8+\delta}$ (OP-Bi2212,
$T_\mathrm{c}\simeq$ 96 K) to investigate the material dependence of magnetic
excitation within the Bi-based superconductor family.

\section{Experimental method}

The high quality OP-Bi2201 and OP-Bi2212 single crystals were grown by the
floating zone method. The hole concentration was optimized by annealing the
samples in N$_2$ flow. The RIXS experiments for OP-Bi2201 collected at 50~K and
200~K were performed with the SAXES instrument \cite{GiacomoSAXES} at the
ADRESS beamline of the Swiss Light Source at the Paul Scherrer Institut
\cite{Strocov2010}, the experimental energy resolution was $\sim150$ meV; RIXS
measurements for OP-Bi2201 and OP-Bi2212 collected at 40~K were performed with
the AXES spectrometer at the beamline ID08 of the European Synchrotron
Radiation Facility (ESRF) \cite{AXES}, the combined energy resolution was
$\sim300$ meV. The x-ray energy was tuned to the maximum of the Cu $L_3$
absorption peak around 931.3 eV. The elastic scattering position was determined
with high accuracy for every momentum transfer by comparing the RIXS spectrum
to that of policrystalline graphite attached on the sample surface. Samples
were cleaved in air some minutes before installation inside the ultrahigh
vacuum measurement chamber ($\sim 3 \times 10^{-9}$~mbar).

The experimental geometry is shown in Fig.~1(a). X-rays are incident on the
sample surface at $\theta$$_i$ and scattered by 2$\theta=130$~deg (constant).
The scattering vector {\bf Q} is denoted using the pseudotetragonal unit cell
with $a=b=3.8$ {\AA} and $c=24.4$ {\AA} for OP-Bi2201, with $a=b=3.86$ {\AA}
and $c=31$ {\AA} for OP-Bi2212, where the axis ${c}$ is normal to the cleaved
sample surface. $\delta$ is the angle between total momentum $\textbf{Q}$ and
sample ${c}$-axis. In the experiment $\delta$ is changed by rotating the sample
around the vertical axis ${b}$ in order to change $Q_{\parallel}$, the
projection of the momentum transfer $\textbf{Q}$ along [100]. Here large
negative $Q_{\parallel}$ corresponds to near grazing-incidence geometry; large
positive $Q_{\parallel}$ corresponds to near grazing-emission geometry. The
x-ray polarization can be chosen parallel ($\pi$) or perpendicular ($\sigma$)
to the horizontal scattering plane. Fig.~1(b) shows the reciprocal space near
the Brillouin zone center. The typical size of the Brillouin zone in cuprates
is 0.81~{\AA}$^{-1}$ (0.5~reciprocal lattice units, rlu) and the maximum of
$Q_{\parallel}$ for 930~eV photons is 0.77~{\AA}$^{-1}$ (0.48~rlu) for
2$\theta=130$~deg. We measured along ($\pm$0.5,0) direction and the thick green
line represents the region explored in this work. We follow previous
conventions \cite{Luciomagnon} and present normalized spectra so that the
integrated intensity of the ${dd}$ excitations ([-3,-1]~eV) equals to one.

\begin{figure}[t]
\begin{center}
\includegraphics[width=0.95\linewidth,angle=0]{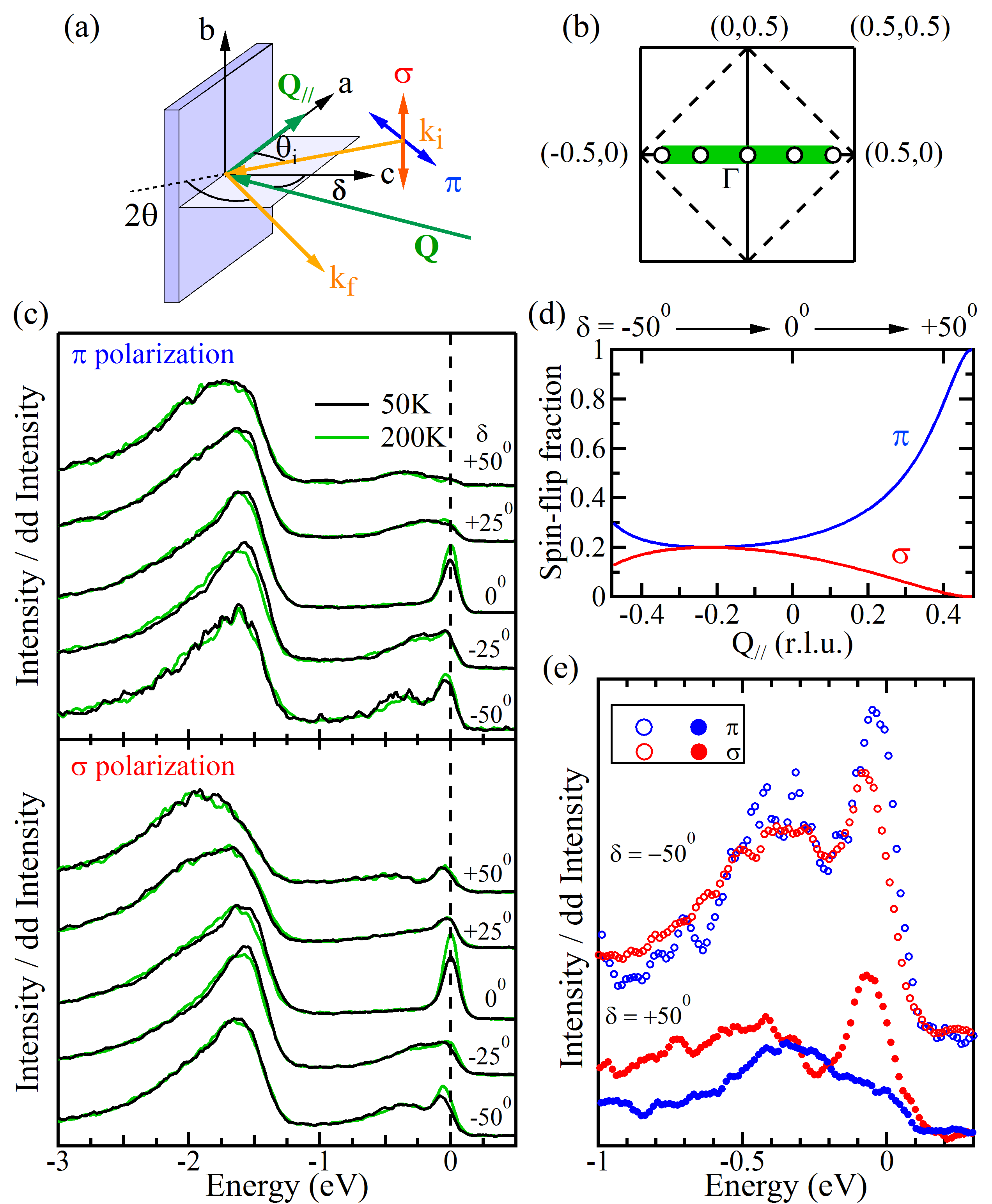}
\end{center}
\caption{(color online) (a) The experimental geometry. (b) Reciprocal-space image,
the nuclear and magnetic first Brillouin zones are drawn with solid and dashed
lines, respectively; the thick green line indicates the range covered by the
experiments. (c) Representative RIXS spectra of OP-Bi2201 for various $\delta$
values in steps of 25~deg as indicated by the black empty circles in panel b, i.e.
from 15~deg grazing incidence to 15~deg grazing emission, as measured
with respect to the sample surface. Top: $\pi$ polarized incident x-rays. Bottom: $\sigma$
polarized incident x-rays. Data were collected at 50~K and 200 K. (d) Single
spin-flip fraction (without orbital excitation) for $\pi$ or $\sigma$ polarizations
according to RIXS cross section calculations for 2$\theta =130$~deg. The
non-spin-flip channel can be spread over elastic, charge excitations, double-spin-flip
excitations and phonons, covering a broad spectral range from 0 to 1 eV energy loss.
This explains why spin-flip channel, less intense but more concentrated in energy,
often the most easily recognizable spectral feature. (e) RIXS spectra measured at
$Q_{\parallel}=-0.4$ rlu with grazing-incidence geometry ($\delta =-50$~deg) and
measured at $Q_{\parallel}=0.4$ rlu with grazing-emission geometry ($\delta=+50$~deg)
for $\pi$- or $\sigma$-polarized light. Data were collected at 50~K. All spectra
are normalized to the integrated intensity of the ${dd}$ excitations.}
\end{figure}

\section{Results}

Fig.~1(c) displays some RIXS spectra at representative $\delta$ angles for
OP-Bi2201, collected at 50~K and 200~K with both $\pi$ and $\sigma$ polarized
incident x-rays. The spectra exhibit, below $-1.5$ eV, ${dd}$ excitations
(transitions of the unpaired hole of Cu$^{2+}$ from the $d_{x^2-y^2}$ to other
$d$ orbitals)\cite{MarcoIOP,ddMateoPRB,ddBrink}, with no obvious temperature
dependence. Two broad peaks around -1.6~eV and around -1.9~eV can be roughly
ascribed to the transitions to the $d_{xy}$ and $d_{xz/yz}$ orbitals, in
analogy to the results obtained in the undoped compounds by Moretti Sala et al.
\cite{MarcoIOP}. The $d_{3z^2-r^2}$ final state, more difficult to discern, is
probably at slightly higher energy loss. The elastic peak is, as usual,
stronger at specular angle ($\delta = 0$~deg) due to the reflectivity from the
surface. The dispersion of the ${dd}$ excitations is as small as that observed
previously in other layered cuprates \cite{MarcoIOP} within the energy
resolution of our experiment.

In layered cuprates the simplest magnetic excitation implies the reversal of
the spin 1/2 at one site, giving origin to a magnon in the antiferromagnetic
parent compounds and to a so-called paramagnon in doped materials. In all cases
the single spin-flip excitation is obtained through the rotation by 90~deg of
the scattering photon polarization vector: it turns out that the only
$\sigma\pi'$ and $\pi\sigma'$ combinations lead to non-zero spin-flip cross
section, where the prime denotes the polarization of the scattered photon
\cite{AmentMagnon,ThomasNcomm}, because due to the $x^2-y^2$ symmetry of the
$3d$ hole, the relevant polarization rotation is that projected on the $ab$
plane. In Fig.~1(d) we show the spin-flip fraction as calculated with the
simplified RIXS cross sections of
Refs.~\onlinecite{AmentMagnon,Luciomagnon,MarcoIOP} for $2 \theta=130$~deg. The
blue (red) line represents the spin-flip fraction of the low energy excitation
spectral weight for incident $\pi$ ($\sigma$) polarization. It must be kept in
mind that, whereas the non-spin-flip part is spread over several contributions
differing in nature (charge excitations, phonons, diffuse elastic, double
spin-flip) and energy, the single spin-flip channel is concentrated in the
resolution limited peak in the antiferromagnetic parent compounds or in the
(often broad) paramagnon in the doped superconductors. This fact makes the
spin-flip excitations more evident and recognizable with respect to the
non-spin-flip excitations. We notice that, at grazing incidence (large negative
$\delta$ values), the single spin-flip fraction is relatively small for both
incident polarizations. This is particularly evident at $\delta=-50$~deg as
shown in Fig.~1(e): for both polarizations the spectra are given by a broad
distribution in the mid-infrared region, provided by a combination of single
and multiple paramagnons, charge and vibrational excitations, all difficult to
disentangle. On the other hand, in a near-grazing-emission geometry (large
positive $\delta$ values), single spin-flip excitations are prominent for $\pi$
and suppressed for $\sigma$ polarization. This is evident at $\delta=+50$~deg
for the peak at $\sim$350~meV in Fig.~1(e). This incident polarization
dependence has been further confirmed by a recent RIXS experiment obtained by
using a soft x-ray polarimeter capable of measuring the polarization of the
scattered radiation\cite{LuciPolarimeter} simultaneously with the spectral
distribution. Therefore we consider the peak at $\sim$350~meV at
$Q_{\parallel}=0.4$ rlu ($\delta=+50$~deg) to be a single spin-flip excitation.

\begin{figure}[t]
\begin{center}
\includegraphics[width=0.95\columnwidth,angle=0]{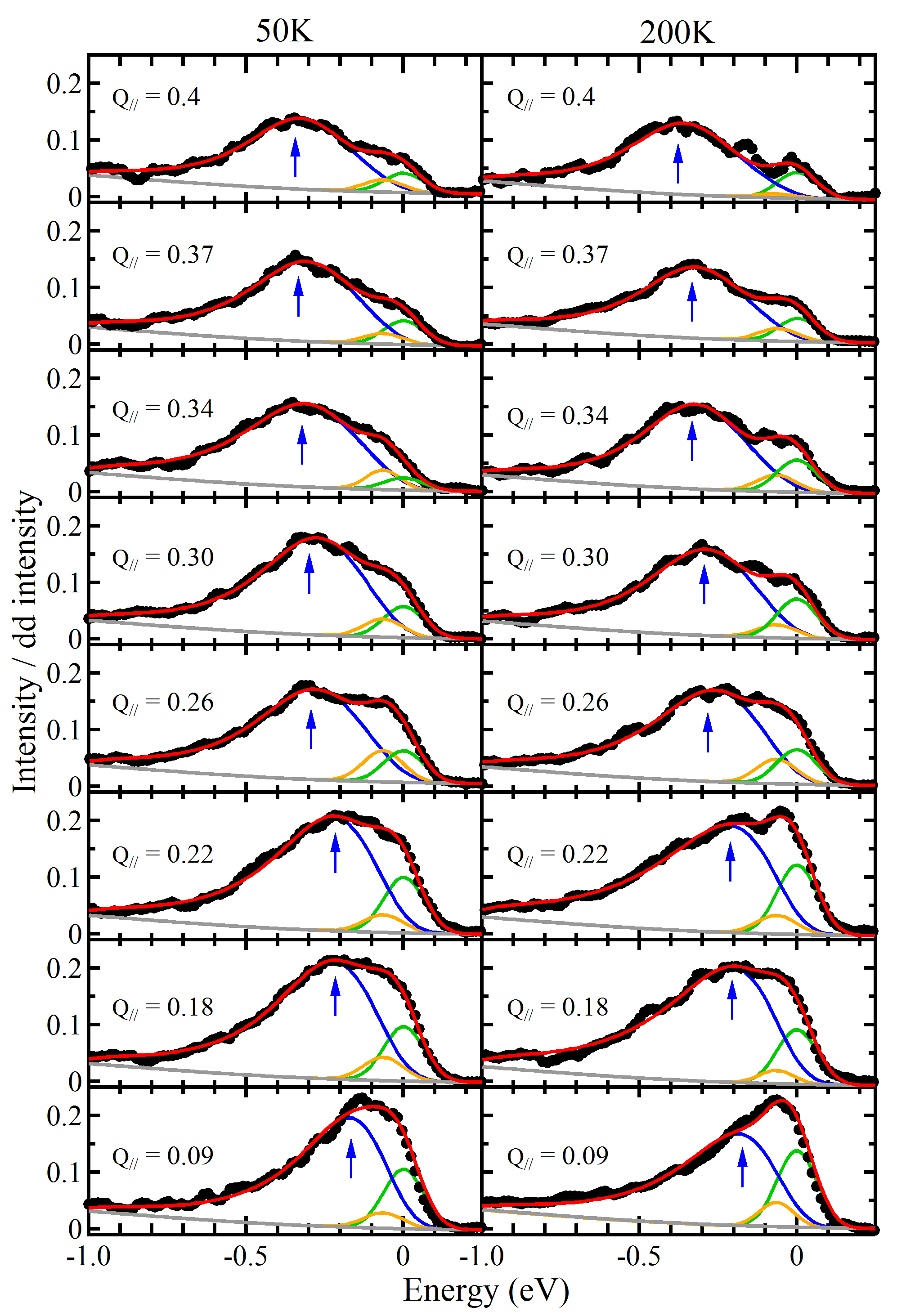}
\end{center}
\caption{(color online) The RIXS spectra of OP-Bi2201 along (0,0)-(0.5,0)
symmetry direction (black solid circles) at 50~K (left) and 200~K (right).
The spectra are decomposed into the dispersive magnetic excitations (blue line),
the elastic scattering (green line), the phonon scattering (orange line), and
the charge- and $dd$-excitations background (gray line). The blue arrow
indicates the paramagnon peak. The data were collected at grazing-emission
using $\pi$ incident polarization. }
\end{figure}

In Fig.~2 we show the evolution of the spin-flip spectral component along the
(0,0)-(0.5,0) symmetry direction. It appears from the raw data that the
paramagnon mode disperses to high energy loss similarly to that of other
superconducting cuprates \cite{LucioLSCO, TaconNatphysics,DeanNatMat,DeanPRL}.
We decompose the spectra into four different contributions
\cite{TaconNatphysics,DeanNatMat}: a resolution-limited Gaussian for the
elastic peak, an anti-symmetrized Lorentzian for the magnetic scattering, a
smooth background for the particle-hole continuum and the tail of $dd$
excitations, and a resolution-limited Gaussian for the dominant phonon which is
a bond stretching longitudinal optical (LO) mode of $\sim$65~meV previously
observed by Raman \cite{Ramanphonon} and high-resolution inelastic x-ray
scattering \cite{Lanzaraphonon}. Its strong coupling to electrons, manifested
in ARPES by a kink in the electronic state dispersion
\cite{Lanzaraphonon,PengCPL}, constitutes a significant part in RIXS. Within
the present experimental accuracy we can not determine the phonon dispersion,
and the crucial implications of its observation in Cu $L_3$ RIXS will be
discussed elsewhere \cite{luciophonon}.

\begin{figure}[t]
\begin{center}
\includegraphics[width=1\columnwidth,angle=0]{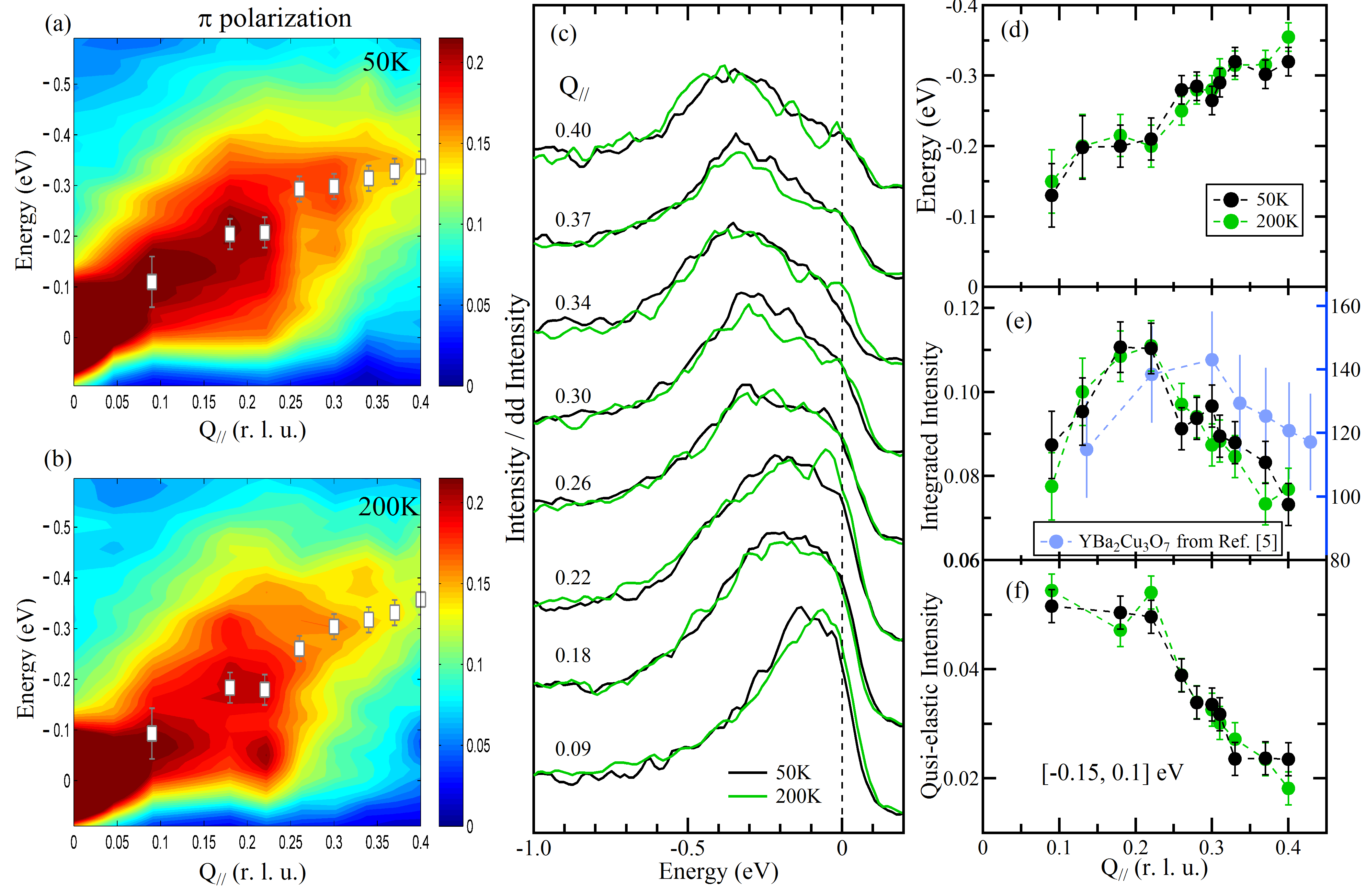}
\end{center}
\caption{(color online) Energy/momentum intensity false-color maps of RIXS
spectra along (0,0)-(0.5,0) symmetry direction measured at (a) 50~K and
(b) 200~K with $\pi$ polarization on OP-Bi2201. The white solid squares
indicate the paramagnon peak positions as determined with the fitting procedure
illustrated in Fig.~2. (c) RIXS spectra at 50~K (black) and 200~K (green) at
selected $Q_{\parallel}$. (d) Experimental paramagnon dispersion and (e) the
integrated intensity of paramagnon peak at 50K (black) and 200K (green)
determined from the fitting procedure. Integrated inelastic intensity of
optimally doped YBa$_2$Cu$_3$O$_7$ from Ref.~\onlinecite{TaconNatphysics}
is superimposed for comparison (blue). (f) Intensity at [-0.15,0.1]~eV energy
window for quasi-elastic signal at 50~K (black) and 200~K (green).
Self-absorption correction has been applied to (e) and (f). The error bars
represent the uncertainty in the fitting. }
\end{figure}

We track the paramagnon peak as denoted by the blue arrows in Fig.~2, and
superimpose it on the energy/momentum intensity map in Fig.~3 for (a) 50~K
(below $T^*$) and (b) 200~K (above $T^*$). From the maps we note that the
intensity of the high energy magnetic excitation shows a decrease above
$Q_{\parallel}\simeq 0.22$ rlu for both temperatures; meanwhile one feature
appears below 100~meV at $Q_{\parallel}\simeq 0.22$ rlu, which will be
discussed subsequently. To visualize if there is any temperature dependence, we
overlap the raw spectra measured at the two temperatures in Fig.~3(c). The two
sets of spectra are almost identical within the experimental uncertainty,
confirming therefore that the effective super-exchange interaction
$J_\textrm{eff}$ is very weakly temperature dependent \cite{INSLCO}.

The energy and integrated intensity of paramagnon determined from the fitting
procedure are compared respectively in Fig.~3(d) and (e) for the two
temperatures. In the antiferromagnetic systems the paramagnon intensity is
expected theoretically to decrease to zero when approaching the $\Gamma$ point
\cite{AmentMagnon,HaverkortLS}. From Fig.~3(e) we observe a similar behavior
below 0.2 rlu also in OP-Bi2201, although the paramagnon intensity could not be
tracked below 0.1 rlu due to the onset, towards $\Gamma$, of the elastic peak
that hinders any possible sub-100 meV inelastic feature with the present
resolution. More interestingly though, the intensity of paramagnon also shows
an abrupt drop above $Q_{\parallel}\simeq 0.22$ rlu. In Fig.~3(f) we integrated
over [-0.15,0.1]~eV energy window to evaluate the quasi-elastic intensity: the
intensity shows a step (or a peak) at $Q_{\parallel}\simeq 0.22$ rlu and then
decreases with momentum for both temperatures. We have applied self-absorption
correction \cite{Matteoarxiv} to the intensity and found that the change of
intensity is less than 10\% near the Brillouin zone boundary, which can not
explain the strong depression of intensity above 0.22 rlu here (nearly 40\%
decrease around the Brillouin zone boundary). For bismuth cuprates the small
correction for the spectra normalized to $dd$ intensity is due to the relative
large pre-edge signal of x-ray absorption spectra, originated from the large
non-resonant absorption of bismuth. The coincident intensity drop of paramagnon
and quasi-elastic signal indicates a nontrivial effect. We notice that for
optimally doped YBa$_2$Cu$_3$O$_7$ in Refs.~\onlinecite{TaconNatphysics}, the
integrated inelastic intensity shows a decrease above 0.3 rlu, a wave vector
characterizing charge order \cite {TaconYBCOarxiv} and bond-buckling phonon
anomaly \cite{Resznikphonon}. Therefore, charge order and phonon anomaly might
play a role in our OP-Bi2201 too, as discussed below. We also notice that
around $Q_{\parallel}\simeq 0.22$ rlu the paramagnon energy in Fig.~3(d) seems
to deviate from a smooth dispersion, which deserves further investigation as
well.

\begin{figure}[t]
\begin{center}
\includegraphics[width=0.95\columnwidth,angle=0]{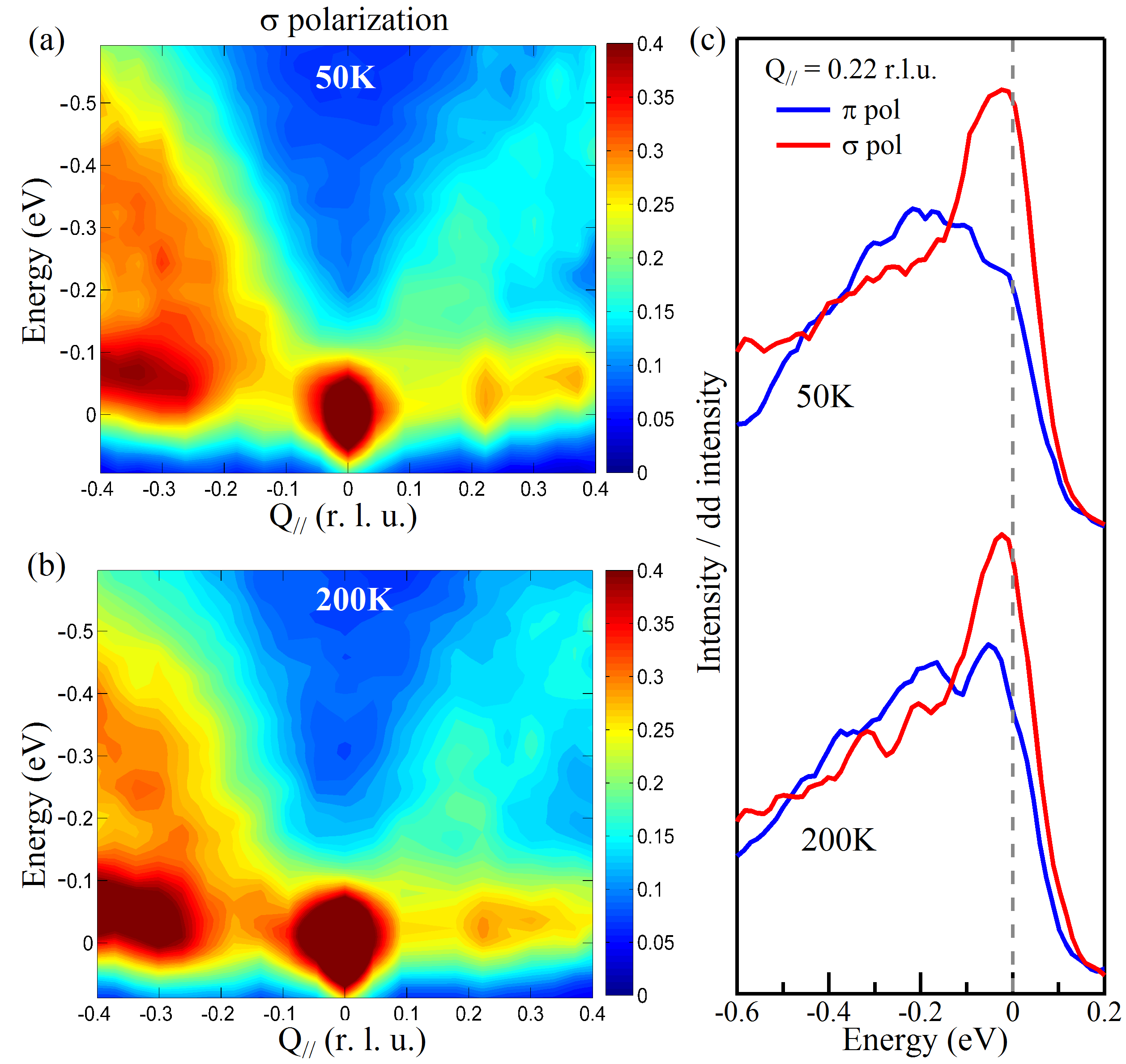}
\end{center}
\caption{(color online) Energy/momentum intensity false-color maps of RIXS
spectra along (-0.5,0)-(0.5,0) symmetry direction measured at (a) 50~K and
(b) 200~K with $\sigma$ polarization on OP-Bi2201. (c) Polarization comparison
of RIXS spectra measured at Q$_{\parallel}=0.22$ rlu at 50~K and 200 K.}
\end{figure}

\section{Discussion}
\subsection{Intensity maps of RIXS spectra}

The concurrent intensity drop of paramagnon and quasi-elastic signal around
Q$_{\parallel}=0.22$ rlu are probably not a fortuitous coincidence. In the map
of Fig.~3(b), we can directly observe a feature below 100~meV at $Q_{\parallel}
\simeq 0.22$ rlu. This can be put in relation with other intriguing phenomena
taking place in Bi2201, such as charge order and strong electron-phonon
coupling. Charge order, with $Q_\textrm{CO}=0.243$ rlu, has been observed
recently in underdoped Bi2201 ($T_\textrm{c}=30$ K) by resonant x-ray
scattering \cite{Damascelliscience} and, with $Q_{\parallel}$$\sim$0.28 rlu, in
OP-Bi2212 by RIXS\cite{HashimotoBi2212}. Here the wave-vector of the low energy
feature, i.e. 0.22 rlu, is compatible with the one of a charge order signal.
However the persistence of the feature up to 200 K, above the presumed charge
order temperature, seems in contrast with the commonly observed temperature
dependence of the charge order in hole-doped cuprates
\cite{GiacomoCDW,Damascelliscience}. In the present RIXS data on OP-Bi2201 we
did not find an evidence for charge order, possibly because of the excessive
disorder, known to be higher in Bi2201 than in Bi2212 \cite{STMBi2201,
STMBi2212}. Unfortunately the lack of an unambiguous evidence of charge order
in the present sample does not allow us to make a direct connection, on that
issue, to the ARPES results, which showed particle-hole symmetry breaking
\cite{Hashimotonphys} and a phase transition below the pseudogap temperature
\cite{Hescience}. On the other hand we can exploit the richness of the RIXS
spectra.

The assignment of the low energy feature can be inspired by observing the RIXS
colormaps of Fig.~4(a) and 4(b), where the $\sigma$ polarization was used. At
positive $Q_{\parallel}$ the $\sigma$ polarization enhances the non-spin-flip
final states, including bi-magnon-like magnetic excitations, particle-hole pair
generation and phonons; at negative $Q_{\parallel}$ the spin-flip and
non-spin-flip excitations have similar intensity. We can unambiguously observe
a peak below 100~meV at $Q_{\parallel} = +0.22$ rlu: this peak is stronger with
$\sigma$ than $\pi$ polarization, as highlighted by the spectra comparison in
Fig.~4(c), a clear demonstration of its non-spin-flip character. Its energy is
fully compatible with a phonon excitation. This assignment is confirmed by the
data at negative $Q_{\parallel}$, where the phonon peak has an almost flat
dispersion but a clear increase in intensity beyond -0.25~rlu. Interestingly,
inelastic x-ray scattering measurement on optimally doped
Bi$_2$Sr$_{1.6}$La$_{0.4}$CuO$_{6+\delta}$ found that the Cu-O bond stretching
(BS) phonon shows a softening and an anomalously broad line-shape around
0.22-0.25 rlu \cite{Lanzaraphonon}. The crossing of two longitudinal phonon
modes and/or the anomalous broadening of the BS phonon may have a relation to
the strong phonon signal around 0.22 rlu in RIXS. Moreover phonon softening and
broadening has been observed around the charge ordering wave-vector in several
copper oxide superconductors \cite{Resznikphonon,Taconphonon,Heydonphonon},
revealing the correlation between charge order and phonon anomalies in
experiments other than RIXS. What RIXS is adding here in the specific case of
OP-Bi2201, is the coincidence of phonon and paramagnon intensity anomalies,
appearing in the putative charge-order wave-vector region. Interestingly
enough, optimally doped YBCO shows a similar maximum of the paramagnon
intensity around its own charge-order wave-vector (0.30 rlu), as shown in
figure 3(e). The q-dependence of the phonon intensity in RIXS and its relation
with charge order goes beyond the scope of this article and will be treated
more systematically elsewhere \cite{luciophonon}. Moreover in the near future
better resolved and more systematic RIXS measurements will help to clarify the
connection of charge- order and phonon anomaly with the paramagnon energy and
intensity evolution.

\begin{figure}[b]
\begin{center}
\includegraphics[width=1\columnwidth,angle=0]{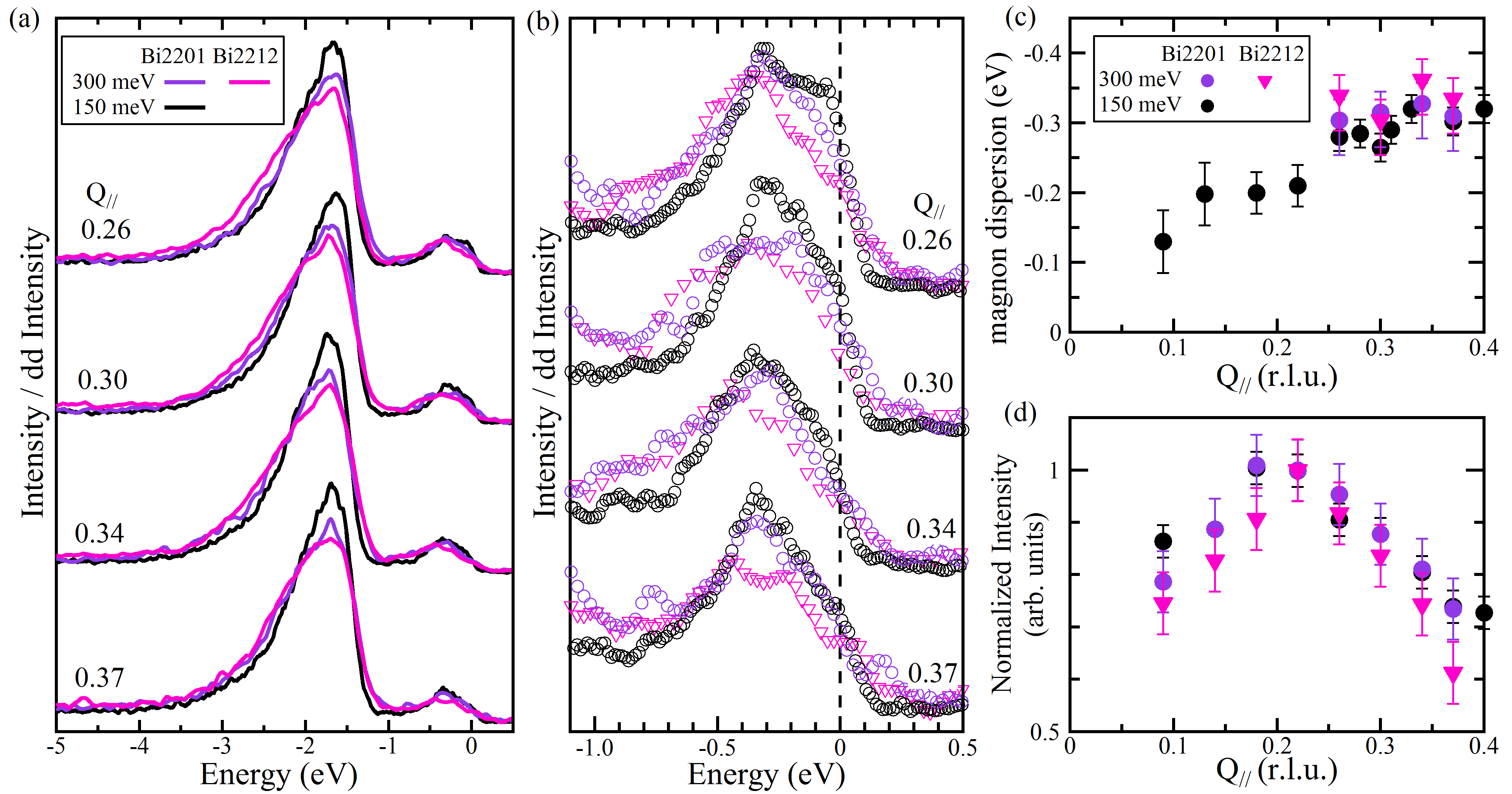}
\end{center}
\caption{(color online) (a) Comparison of RIXS spectra of OP-Bi2201 and OP-Bi2212
at large $Q_{\parallel}$ measured with $\pi$ polarization. Two data sets were
collected for OP-Bi2201, at 50~K and 40~K with energy resolution of 150~meV
(black) and 300~meV (purple) respectively; OP-Bi2212 data were collected at
40~K with energy resolution of 300~meV (magenta). (b) Enlarged view of the
low energy portion. (c) The paramagnon energies of OP-Bi2201 and OP-Bi2212
from the fitting of the spectra with experimental resolution of 300~meV are
overlaid with paramagnon energy of OP-Bi2201 at 50~K reproduced from Fig. 3(d).
The large error bars are due to the fitting uncertainty for the lower resolution
spectra. (d) Integrated intensities of RIXS spectra with energy window [-0.8,0] eV,
normalized to the respective value at $Q_{\parallel}=$0.22 rlu. Self-absorption
correction has been applied to the intensities. The error bars represent the
uncertainty in determining the spectral weight. }
\end{figure}

\subsection{Relation between $J_\textrm{eff}$ and $T_\textrm{c}$}

In Fig.~5 we compare the RIXS spectra of OP-Bi2201 and OP-Bi2212 at large
$Q_{\parallel}$ for $\pi$ polarization. In panel (a), the two data sets for
OP-Bi2201 (one at 50~K with 150~meV resolution, one at 40~K with 300~meV
resolution) are compatible with each other, once the difference in resolution
is taken into account. For OP-Bi2212 (40 K, 300~meV resolution) we observe that
the $dd$ multiplet is more extended towards higher energy, due to the
$d_{3z^2-r^2}$ final state being more separated from the $d_{x^2-y^2}$, due to
the absence of one apical oxygen in the bilayer compound with respect to the
single layer Bi2201. Panel (b) shows an enlarged view of the low energy
portion. The paramagnon energies are similar in the two samples. Following the
usual fitting procedure\cite{DeanNatMat,TaconNatphysics}, we obtain that the
paramagnon energy at zone boundary is $\sim$350 meV, as shown in Fig.~5(c). At
small $Q_\parallel$ the fitting for spectra with lower resolution is too
uncertain and we cannot plot the corresponding points in panel (c). We present
the integrated intensities for OP-Bi2201 and OP-Bi2212 in Fig.~5(d), which are
very similar with a consistent drop above $\sim$0.22 rlu, suggesting a common
interplay in Bi-based cuprate family.

These results are in stark contrast with those recently published by Dean et
al. \cite{Deanarxiv}, who have found that the paramagnon energy at zone
boundary is substantially higher in Bi$_2$Sr$_2$Ca$_2$Cu$_3$O$_{10+\delta}$
(Bi-2223, $T_\textrm{c}=109$ K) than in Bi$_{2+x}$Sr$_{2-x}$CuO$_{6+\delta}$
(Bi-2201, $T_\textrm{c} \simeq 1$ K). In Ref.~\onlinecite{Deanarxiv} the
paramagnon energies are assigned at 295~meV for Bi-2201 and 347~meV for Bi-2223
close to $(1/2,0,L)$. From this result the authors argued that $T_\textrm{c}$
scales monotonically with $J_\textrm{eff}$. However, previous results in Bi2212
had rather shown an opposite trend, i.e., the softening with doping of the
magnetic excitation energy, namely from $\sim350$~meV in the heavily underdoped
nonsuperconducting sample to $\sim300$~meV in the optimally doped one
($T_\textrm{c}=92$ K) \cite{DeanPRL}. Here we find that the paramagnon energies
of OP-Bi2201 ($T_\mathrm{c}\simeq34$ K) and OP-Bi2212 ($T_\mathrm{c}\simeq96$
K) are indeed similar to that of Bi-2223 ($T_\textrm{c}=$109 K) and of heavily
underdoped Bi2212. Therefore our results confirm for the Bi22$nm$ family what
was already known for YBCO and LSCO
\cite{TaconNatphysics,TaconPRBOD,DeanNatMat}, that the value of the magnetic
excitation energy at zone boundary does not correlate directly to
$T_\textrm{c}$. The exception found in Ref.~\onlinecite{Deanarxiv} is possibly
due to the different local structure (Cu-O distance, Cu-O-Cu bond angle) of the
Bi-doped compound, chosen for its especially low $T_\textrm{c}$ at optimal
doping. Although there is a good probability that magnetic fluctuations play a
decisive role in high-$T_\textrm{c}$ superconductivity, the maximum of
$T_\textrm{c}$ for different materials may be more strongly influenced by other
factors than just the value of the superexchange coupling.

\section{Conclusions}

In the RIXS spectra measured with the appropriate conditions ($\pi$
polarization, positive $Q_\parallel$ values), we have observed the simultaneous
intensity decrease of the paramagnon and the quasi-elastic signal above
$\sim$0.22 rlu, the wave-vector of the putative charge-order in OP-Bi2201. The
tiny temperature dependence and the evidence of a concurrent onset, with
$\sigma$ polarization, of the phonon signal at that special wave-vector, hint
at a combined effect, on the magnetic excitation spectrum, of electron-phonon
coupling and incipient charge-order. Further insight into this issue could come
from considering the influence of phonon and charge order on the spin dynamical
structure factor. Moreover we give here a further confirmation of the
robustness of magnetic excitations across the phase diagram of high
$T_\textrm{c}$ cuprate superconductors. There is, however, an increasing
general evidence that, in cuprates, spin excitations get coupled, via the
electron-phonon interaction, to both lattice modes and charge order, therefore
providing a ubiquitous ingredient for the superconductivity pairing mechanisms.
A better clarification of this three-actors scenario (spin excitations,
electron-phonon coupling, charge order) will require further systematic use of
high resolution resonant elastic and inelastic x-ray scattering.

\begin{acknowledgments}
The spectra at the ADRESS beam line of the Paul Scherrer Institut were measured
using the SAXES instrument developed jointly by Politecnico di Milano, SLS, and
EPFL. This work is supported by the PIK project POLARIX of the Italian Ministry
of Research (MIUR).
\end{acknowledgments}


\begin{thebibliography}{99}

\bibitem{AmentMagnon} L. J. P. Ament, G. Ghiringhelli, M. M. Sala, L.
    Braicovich, and J. van den Brink, Phys. Rev. Lett. 103, 117003 (2009).
\bibitem{HaverkortLS} M. W. Haverkort, Phys. Rev. Lett. 105, 167404 (2010).
\bibitem{IshiElectrondope} K. Ishii, M. Fujita, T. Sasaki, M. Minola, G.
    Dellea, C. Mazzoli, K. Kummer, G. Ghiringhelli, L. Braicovich, T. Tohyama,
    K. Tsutsumi, K. Sato, R. Kajimoto, K. Ikeuchi, K. Yamada, M. Yoshida, M.
    Kurooka and J. Mizuki, Nat. Commun. 5, 3714 (2014).
\bibitem{LucioLSCO} L. Braicovich, J. van den Brink, V. Bisogni, M. M. Sala, L.
    J. P. Ament, N. B. Brookes, G. M. De Luca, M. Salluzzo, T.
    Schmitt, V. N. Strocov, and G. Ghiringhelli, Phys. Rev. Lett. 104, 077002
    (2010).
\bibitem{TaconNatphysics} M. Le Tacon, G. Ghiringhelli, J. Chaloupka, M. M.
    Sala, V. Hinkov, M.W. Haverkort, M. Minola, M. Bakr, K. J. Zhou, S.
    Blanco-Canosa, C. Monney, Y. T. Song, G. L. Sun, C. T. Lin, G. M. De Luca,
    M. Salluzzo, G. Khaliullin, T. Schmitt, L. Braicovich, and B. Keimer, Nat.
    Phys. 7, 725 (2011).
\bibitem{DeanNatMat} M. P. M. Dean, G. Dellea, R. S. Springell, F.
    Yakhou-Harris, K. Kummer, N. B. Brookes, X. Liu, Y-J. Sun, J. Strle, T.
    Schmitt, L. Braicovich, G. Ghiringhelli, I. Bozovic and J. P. Hill, Nat.
    Mater. 12, 1019 (2013).
\bibitem{DeanPRL} M. P. M. Dean, A. J. A. James, R. S. Springell, X. Liu, C.
    Monney, K. J. Zhou, R. M. Konik, J. S. Wen, Z. J. Xu, G. D. Gu, V. N.
    Strocov, T. Schmitt, and J. P. Hill, Phys. Rev. Lett. 110, 147001 (2013).
\bibitem{TaconPRBOD} M. Le Tacon, M. Minola, D. C. Peets, M. Moretti Sala, S.
    Blanco-Canosa, V. Hinkov, R. Liang, D. A. Bonn, W. N. Hardy, C. T. Lin, T.
    Schmitt, L. Braicovich, G. Ghiringhelli, and B. Keimer, Phys. Rev. B 88,
    020501 (2013).
\bibitem{TranquadaStripe} J. M. Tranquada, B. J. Sternlieb, J. D. Axe, Y.
    Nakamura and S. Uchida, Nature (London) 375, 561 (1995).
\bibitem{FujitaStripe} M. Fujita, H. Goka, K. Yamada, and M. Matsuda, Phys.
    Rev. Lett. 88, 167008 (2002).
\bibitem{Stripexray} P. Abbamonte, A. Rusydi, S. Smadici, G. D. Gu, G. A.
    Sawatzky, and D. L. Feng, Nat. Phys. 1, 155 (2005).
\bibitem{JulienNMR} T. Wu, H. Mayaffre, S. Kr\"amer, M. Horvatic, C. Berthier,
    W. N. Hardy, R. X. Liang, D. A. Bonn, and M.-H. Julien, Nature (London)
    477, 191 (2011)
\bibitem{GiacomoCDW} G. Ghiringhelli, M. Le Tacon, M. Minola, S. Blanco-Canosa,
    C. Mazzoli, N. B. Brookes, G. M. De Luca, A. Frano, D. G. Hawthorn, F. He,
    T. Loew,M.M. Sala, D. C. Peets, M. Salluzzo, E. Schierle, R. Sutarto, G. A.
    Sawatzky, E. Weschke, B. Keimer, and L. Braicovich, Science 337, 821
    (2012).
\bibitem{REXSCDW} A. J. Achkar, R. Sutarto, X. Mao, F. He, A. Frano, S.
    Blanco-Canosa, M. Le Tacon, G. Ghiringhelli, L. Braicovich, M. Minola, M.
    Moretti Sala, C. Mazzoli, R. Liang, D. A. Bonn,W. N. Hardy, B. Keimer, G.
    A. Sawatzky, and D. G. Hawthorn, Phys. Rev. Lett. 109, 167001 (2012).
\bibitem{changCDW} J. Chang, E. Blackburn, A. T. Holmes, N. B. Christensen, J.
    Larsen, J. Mesot, R. Liang, D. A. Bonn, W. N. Hardy, A. Watenphul, M. v.
    Zimmermann, E. M. Forgan, and S. M. Hayden, Nat. Phys. 8, 871 (2012).
\bibitem{diffractionCDW} E. Blackburn, J. Chang, M. Hucker, A. T. Holmes, N. B.
    Christensen, R. Liang, D. A. Bonn, W. N. Hardy, U. Rutt, O. Gutowski, M. v.
    Zimmermann, E. M. Forgan, and S. M. Hayden, Phys. Rev. Lett. 110, 137004
    (2013).
\bibitem{Damascelliscience} R. Comin, A. Frano, M. M. Yee, Y. Yoshida, H.
    Eisaki, E. Schierle, E. Weschke, R. Sutarto, F. He, A. Soumyanarayanan, Y.
    He, M. Le Tacon, I. Elfimov, J. E. Hoffman, G. Sawatzky, B. Keimer, and A.
    Damascelli, Science 343, 390 (2014).
\bibitem{STMBi2212} E. H. da Silva Neto, P. Aynajian, A. Frano, R. Comin, E.
    Schierle, E. Weschke, A. Gyenis, J. S. Wen, J. Schneeloch, Z. J. Xu, S.
    Ono, G. D. Gu, M. Le Tacon, and Ali Yazdani, Science 343, 393 (2014).
\bibitem{HashimotoBi2212} M. Hashimoto, G. Ghiringhelli, W.-S. Lee, G. Dellea,
    A. Amorese, C. Mazzoli, K. Kummer, N. B. Brookes, B. Moritz, Y. Yoshida, H.
    Eisaki, Z. Hussain, T. P. Devereaux, Z.-X. Shen, and L. Braicovich, Phys.
    Rev. B 89, 220511(R) (2014).
\bibitem{TaconYBCOarxiv} S. Blanco-Canosa, A. Frano, E. Schierle, J. Porras, T.
    Loew, M. Minola, M. Bluschke, E. Weschke, B. Keimer, M. Le Tacon, Phys.
    Rev. B 90, 054513 (2014).
\bibitem{CominNBCO} E. H. da Silva Neto, R. Comin, F. He, R. Sutarto, Y. Jiang,
    R. L. Greene, G. A. Sawatzky, and A. Damascelli, Science 347, 6219 (2015).
\bibitem{Hashimotonphys} M. Hashimoto, R.-H. He, K. Tanaka, J.-P. Testaud, W.
    Meevasana, R. G. Moore, D. H. Lu, H. Yao, Y. Yoshida, H. Eisaki, T. P.
    Devereaux, Z. Hussain and Z.-X. Shen, Nat. Phys. 6, 414 (2010).
\bibitem{Hescience} R.-H. He, M. Hashimoto, H. Karapetyan, J. D. Koralek, J. P.
    Hinton, J. P. Testaud, V. Nathan, Y. Yoshida, Hong Yao, K. Tanaka, W.
    Meevasana, R. G. Moore, D. H. Lu, S.-K. Mo, M. Ishikado, H. Eisaki, Z.
    Hussain, T. P. Devereaux, S. A. Kivelson, J. Orenstein, A. Kapitulnik, and
    Z.-X. Shen, Science 331, 1579 (2011).
\bibitem{GiacomoSAXES} G. Ghiringhelli, A. Piazzalunga, C. Dallera, G. Trezzi,
    L. Braicovich, T. Schmitt, V. N. Strocov, R. Betemps, L. Patthey, X. Wang,
    and M. Grioni, Rev. Sci. Instrum. 77, 113108 (2006).
\bibitem{Strocov2010} V. N. Strocov, T. Schmitt, U. Flechsig, T. Schmidt, A.
    Imhof, Q. Chen, J. Raabe, R. Betemps, D. Zimoch, J. Krempasky, X. Wang, M.
    Grioni, A. Piazzalunga and L. Patthey, J. Synchrotron Radiat. 17, 631
    (2010).
\bibitem{AXES} C. Dallera, E. Puppin, G. Trezzi, N. Incorvaia, A. Fasana, L.
    Braicovich, N. B. Brookes, and J. B. Goedkoop, J. Synchrotron Radiat. 3,
    231 (1996).
\bibitem{Luciomagnon} L. Braicovich, M. Moretti Sala, L. J. P. Ament, V.
    Bisogni, M. Minola, G. Balestrino, D. Di Castro, G. M. De Luca, M.
    Salluzzo, G. Ghiringhelli, and J. van den Brink, Phys. Rev. B 81, 174533
    (2010).
\bibitem{MarcoIOP} M. Moretti Sala, V. Bisogni, C. Aruta, G. Balestrino, H.
    Berger, N. B. Brookes, G. M. de Luca, D. Di Castro, M. Grioni, M. Guarise,
    P. G. Medaglia, F. Miletto Granozio, M. Minola, P. Perna, M. Radovic, M.
    Salluzzo, T. Schmitt, K. J. Zhou, L. Braicovich and G. Ghiringhelli, New
    Journal of Phys. 13, 043026 (2011).
\bibitem{ddMateoPRB} M. Minola, D. Di Castro, L. Braicovich, N. B. Brookes, D.
    Innocenti, M. Moretti Sala, A. Tebano, G. Balestrino, and G. Ghiringhelli,
    Phys. Rev. B 85, 235138 (2012).
\bibitem{ddBrink} L. Hozoi, L. Siurakshina1, P. Fulde and J. van den Brink,
    Sci. Rep. 1, 65 (2011).
\bibitem{ThomasNcomm} C. J. Jia, E. A. Nowadnick, K. Wohlfeld, Y. F. Kung,
    C.-C. Chen, S. Johnston, T. Tohyama, B. Moritz, and T. P. Devereaux, Nat.
    Commun. 5, 4314 (2014).
\bibitem{LuciPolarimeter} L. Braicovich, M. Minola, G. Dellea, M. Le Tacon, M.
    Moretti Sala, C. Morawe, J-Ch. Peffen, R. Supruangnet, F. Yakhou, G.
    Ghiringhelli, and N. B. Brookes, Rev. Sci. Instrum. \textbf{85}, 115104
    (2014).
\bibitem{Ramanphonon} S. Sugai, H. Suzuki, Y. Takayanagi, T. Hosokawa, and N.
    Hayamizu, Phys. Rev. B 68, 184504 (2003).
\bibitem{Lanzaraphonon} J. Graf, M. d'Astuto, C. Jozwiak, D. R. Garcia, N. L.
    Saini, M. Krisch, K. Ikeuchi, A. Q. R. Baron, H. Eisaki, and A. Lanzara,
    Phys. Rev. Lett. 100, 227002 (2008).
\bibitem{PengCPL}Y. Y. Peng, J. Q. Meng, L. Zhao, Y. Liu, J. F. He, G. D. Liu,
    X. L. Dong, S. L. He, J. Zhang, C. T. Chen, Z. Y. Xu and X. J. Zhou, Chin.
    Phys. Lett. 30, 067402 (2013).
\bibitem{luciophonon} L. Braicovich, et al. unpublished.
\bibitem{INSLCO} R. Coldea, S. M. Hayden, G. Aeppli, T. G. Perring, C. D.
    Frost, T. E. Mason, S.-W. Cheong, and Z. Fisk et al., Phys. Rev. Lett. 86,
    5377 (2001).
\bibitem{Matteoarxiv} M. Minola, G. Dellea, H. Gretarsson, Y. Y. Peng, Y. Lu,
    J. Porras, T. Loew, F. Yakhou, N. B. Brookes, Y. B. Huang, J. Pelliciari,
    T. Schmitt, G. Ghiringhelli, B. Keimer, L. Braicovich and M. Le Tacon,
    arXiv:1502.02583 (2015).
\bibitem{Resznikphonon} D. Reznik, Physica C 481, 75-92 (2012).
\bibitem{STMBi2201} W. D. Wise, K. Chatterjee, M. C. Boyer, T. Kondo, T.
    Takeuchi, H. Ikuta, Z. J. Xu, J. S. Wen, G. D. Gu, Y. Wang, and E. W.
    Hudson, Nat. Phys. 5, 213 (2009).
\bibitem{Taconphonon} M. Le Tacon, A. Bosak, S. M. Souliou, G. Dellea, T. Loew,
    R. Heid, K-P. Bohnen, G. Ghiringhelli, M. Krisch and B. Keimer, Nat. Phys.
    10, 52 (2013).
\bibitem{Heydonphonon} E. Blackburn, J. Chang, A. H. Said, B. M. Leu, Ruixing
    Liang, D. A. Bonn, W. N. Hardy, E. M. Forgan, and S. M. Hayden, Phys. Rev.
    B 88, 054506 (2013).
\bibitem{Deanarxiv} M. P. M. Dean, A. J. A. James, A. C. Walters, V. Bisogni,
    I. Jarrige, M. Hucker, E. Giannini, M. Fujita, J. Pelliciari, Y. B. Huang, R.
    M. Konik, T. Schmitt, and J. P. Hill, Phys. Rev. B 90, 220506 (2014).



\end{thebibliography}
\end{document}